# A unified theory of chaos linking nonlinear dynamics and statistical physics

Chi-Sang Poon[1][1], Cheng Li[1,2], and Guo-Qiang Wu[2]

[1]Harvard-MIT Division of Health Sciences and Technology, Massachusetts Institute of Technology, Cambridge, MA 02139, USA
[2]Department of Mechanics and Engineering Science, Fudan University, Shanghai 200433, People's Republic of China

A fundamental issue in nonlinear dynamics and statistical physics is how to distinguish chaotic from stochastic fluctuations in short experimental recordings. This dilemma underlies many complex systems models from stochastic gene expression or stock exchange to quantum chaos. Traditionally, deterministic chaos is characterized by "sensitive dependence on initial conditions" as indicated by a positive Lyapunov exponent. However, ambiguity arises when applying this criterion to real-world data that are corrupted by measurement noise or perturbed nonautonomously by exogenous deterministic or stochastic inputs. Here, we show that a positive Lyapunov exponent is surprisingly neither necessary nor sufficient proof of deterministic chaos, and that a nonlinear dynamical system under deterministic or stochastic forcing may exhibit multiple forms of nonautonomous chaos assessable by a noise titration assay. These findings lay the foundation for reliable analysis of low-dimensional chaos for complex systems modeling and prediction of a wide variety of physical, biological, and socioeconomic data.

Detection of deterministic chaos in complex time series is an open problem in many branches of physics[1-4], econometrics[5-8] and biology and medicine[9-12]. The crux of the conundrum is that experimental data are inevitably corrupted by measurement and/or dynamic noise, which may cause false detection[13,14] or induce chaos-like fluctuations in nonchaotic systems[15-18]. In quantum chaos, for instance, measurement noise due to the uncertainty principle precludes direct assessment of the long-sought quantum equivalents of classical chaos[4]. These considerations have rekindled interests in stochastic vis-à-vis deterministic approaches to studies of dynamical complexity in many fields[12,19-21]. A fundamental issue is that, whereas the notion of noise has taken hold in statistical physics since Brown-Einstein-Langevin-Perrin[22], a precise definition of low-dimensional chaos has yet to reach consensus in the literature. Beginning with Poincaré and Lorenz[23] it has been widely accepted that a hallmark of deterministic chaos is "sensitive dependence on initial conditions", as indicated by a positive largest Lyapunov exponent (LLE > 0)[24]. A more rigorous mathematical definition is due to Li and Yorke[25], who first coined the term "chaos" in physics. Together, these distinct definitions of deterministic chaos lay the foundation of modern nonlinear dynamics theory although their supposed equivalency remains unproven.

In what follows, we show that contrary to general belief, a positive LLE is neither necessary nor sufficient proof of deterministic chaos, and that the widespread adoption of LLE > 0 as a gold standard for chaos is the root of many controversies and ambiguities in the literature. Rather, a nonlinear low-dimensional dynamical system under deterministic or stochastic forcing may exhibit multiple forms of nonautonomous chaos assessable by a novel noise titration assay. These observations lead to a powerful test of low-dimensional chaos under measurement and dynamic noise that is applicable to the modeling

---
[1] Correspondence: Dr. Chi-Sang Poon
Email: cpoon@mit.edu
Tel: +1 617-258-5405; Fax: +1 617-258-7906

and prediction of a wide range of physical, biological, and socioeconomic data.

## Distinguishing linear and nonlinear dynamics

Reliable assay of complex time series should as a first step meet the following basic criteria: (*i*) *specificity* to discriminate linear and nonlinear correlations (dependences); (*ii*) *sensitivity* to analyze short datasets; (*iii*) *robustness* against linear filtering of data; (*iv*) *robustness* against white or linearly correlated measurement noise. Criteria (*ii-iv*) are necessary because experimental data are typically short, noisy and nonstationary, and often subjected to signal conditioning.

A popular approach to distinguish linear and nonlinear correlations in short time series (criteria *i-ii*) is the surrogate data method via statistical bootstrapping [26,27]. In its most general form[28] the surrogate data test is equivalent to a test of time irreversibility[29], an equivalence relation that often goes unnoticed in the literature. In ref.[27] it is shown that a time-reversible signal may become time-irreversible and fail the surrogate data test after linear filtering (criterion *iii*). Neither the surrogate data nor time reversibility model allows for measurement noise (criterion *iv*). This presents a dilemma since measurement noise must be filtered out before the surrogate data test can be applied. As cautioned by Ruelle[30], experimental data that are corrupted by significant measurement noise or subjected to signal filtering (i.e., not 'raw') may be vulnerable to false positive detections of nonlinearity.

Another well-known approach to distinguish linear and nonlinear correlations is the nonlinear forecasting method[9]. In this method, nonlinearity is detected by comparing the decay of short-term forecasts (n-step-ahead predictions) by a nonlinear and a linear model. Although this elegant approach resolves many of the above issues (criteria *i-iii*) when measurement noise is white, it is prone to false negatives when measurement noise is linearly correlated (criterion *iv*) or when the deterministic chaos has a strong pseudo-periodic background[31].

Furthermore, the assumption of exponential time-dependent decay of nonlinear predictability is predicated upon the condition LLE > 0 in the absence of measurement noise as a signature of chaos. As will be seen below, neither of these assumptions are necessary conditions for low-dimensional chaos.

These difficulties are circumvented by the Volterra autoregressive (VAR) model for nonlinear detection (see Methods). By fitting (instead of forecasting) the data, the VAR method is more robust to attendant linearly correlated noise or pseudo-periodic fluctuations than nonlinear forecasting[32]. Importantly, this method forms the basis for the following 'litmus test' of chaos independent of the LLE.

## Titration of deterministic chaos in measurement noise

### *Autonomous deterministic chaos*

Current methods of estimating the LLE require long datasets[13] and are unreliable under sizable measurement noise[33]. These fundamental difficulties are rectified by the noise titration method (see Methods), which provides a sufficient test of chaos and a quantitative measure of relative chaos intensity in short, noisy data[34]. We call this a 'litmus test' because it literally works like a chemical titration of acid (chaos) against alkaline (added white or linearly correlated noise) with a pH-sensitive litmus agent (VAR model). The mathematical underpinning of this technique is that nonchaotic dynamics such as nonlinear oscillations can be well described by linear models with equivalent Fourier spectra especially in the presence of measurement noise, whereas linearly correlated processes are also best represented by linear models. Thus, only truly chaotic dynamics cannot be described by linear models and must be represented by nonlinear models. By adding random noise to titrate chaos, this procedure is inherently robust to (white or linearly correlated) measurement noise. Indeed, since experimental data is inevitably 'auto-titrated' by measurement noise, the detection of VAR nonlinearity alone constitutes a sufficient proof of chaos in noisy data[34,35]. This auto-titration property has been used to advantage to provide an estimate of the (temporal) average chaos intensity in terms of the detection rate (DR), defined as the percentage of nonlinearity detection in a histogram of consecutive short (<1,000 points) data segments[35,36].

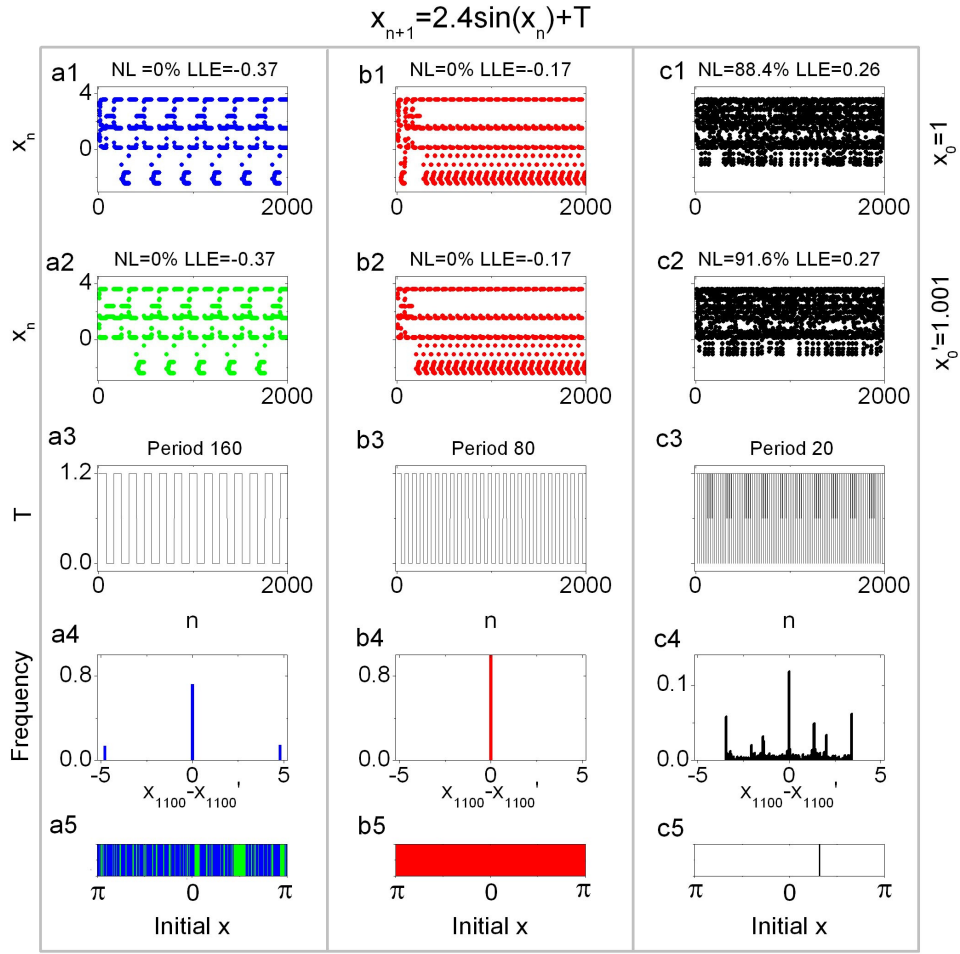

Fig. 1 Distinct forms of induced nonlinear dynamics of the nonautonomous system in Eq. 1: (a) 2:1 entrained oscillations with two antiphase orbits (180o phase difference) demonstrating discrete-event chaos; (b) nonchaotic 1:1 entrained oscillations; and (c) nonautonomous deterministic chaos with aperiodic fluctuations. The results in a-c are for three different square waves T(t) with closely-spaced initial conditions: x0 = 1.000 (a1, b1, c1) and x0' = 1.001 (a2, b2, c2). T(t) switches between 0 and 1.2 with a period of 160 (a3), 80 (b3), and 20 (c3); the corresponding duty cycles are 50%, 50% and 95% (i.e., resetting to 0 every 20 steps). a4, b4, c4 are histograms of the difference (x1100 - x1100') resulting from two initial conditions x0 and x0', where x0 is uniformly random on (-π, π) and x0'=x0+0.001. a5, b5, c5 are basins of attraction. For a5, the initial conditions on (−π, π) that lead to the antiphase solutions in a1 and a2 are marked in blue and green respectively demonstrating fractal basin boundaries. For b5, all initial conditions on (−π, π) lead to identical orbits shown in b1 and b2. For c5, the map is driven to deterministic chaos starting from any initial conditions, and the representative initial condition x0=1 is plotted. NL, noise limit; LLE, largest Lyapunov exponent calculated directly from the equation.

## Nonautonomous deterministic chaos

Traditionally, 'deterministic chaos' is generally taken as aperiodic fluctuations resulting from a low-dimensional nonlinear dynamical system that is 'autonomous' (i.e., with no deterministic or dynamic noise inputs). However, deterministic chaos may also occur when a nonchaotic nonlinear system is driven by deterministic inputs. We call this 'nonautonomous deterministic chaos' (Table 1). Classic examples are the periodically-driven van der Pol oscillator[37,38] and damped pendulum[39], seasonally-forced epidemic model[40], and the quantum kicked top[4].

Of note, deterministic chaos does not necessarily require signal aperiodicity. A well-known example of nonaperiodic chaos with sensitive dependence on initial conditions is the
3

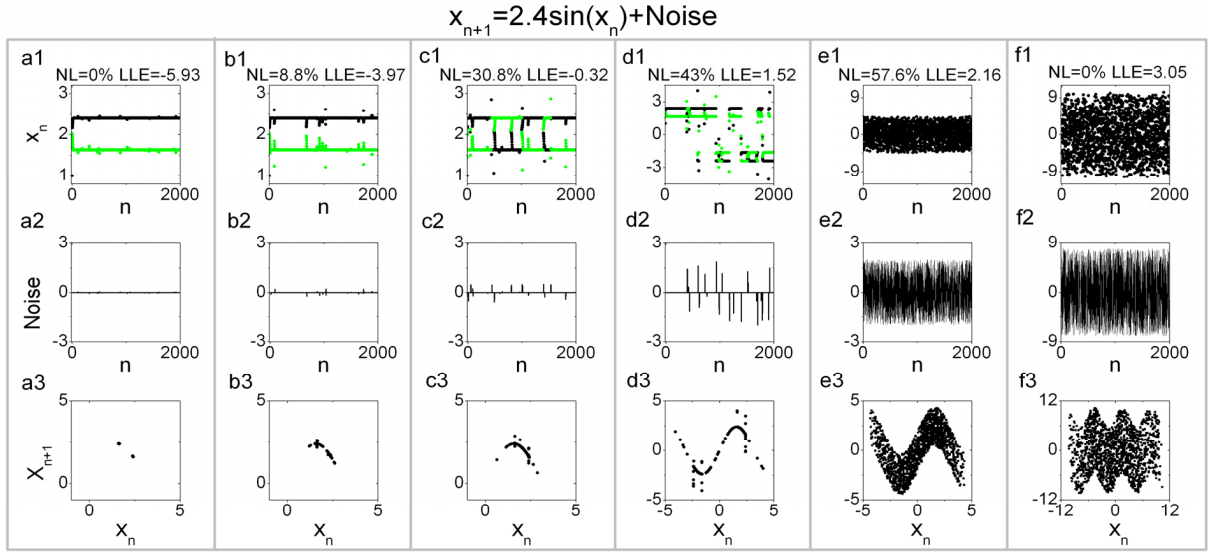

Fig. 2 Noise-induced chaos and noise annihilation of chaos. The nonautonomous system (Eq. 2) is perturbed by increasing levels of dynamic noise (a2-f2) of the form $Y_n \cdot \eta_n \cdot A$ where $Y_n = 1$ or $0$ with probabilities $q$ or $1-q$ respectively; $\eta_n$ is a random variable with uniform distribution between [-2 2]; and A is an amplification factor. In a2-f2, $q = 0.01, 0.01, 0.01, 0.01, 1, 1$; $A = 0.03, 0.15, 0.3, 1, 1$, and $4$ respectively. a1-f1 show the resulting time series with initial condition $x_0=1.000$ and a3-f3 show corresponding first return maps. a1 and a3: the orbit is a period-2 oscillation with distinct phases of the orbit labeled black and green (time sequence: black\ to green \to black…..). b1 and b3: with increased dynamic noise, the period-2 orbit becomes fractal. c1 and c3: with even greater dynamic noise the orbit undergoes intermittent phase resetting, as indicated by the intermittent reversal of the black and green time sequences. d1 and d3: the orbit undergoes both phase resetting and basin hopping, as indicated by the intermittent quantum jumps between the positive and negative orbits. e1 and e3: the orbit becomes aperiodic. Noise titration (with the Mann-Whitney-Wilcoxon test for non-Gaussian residual errors for VAR nonlinear and linear modeling) shows noise-induced chaos in b1-e1 where the orbits morph into chimeric deterministic-stochastic states; and noise annihilation of chaos in f1, where the noise-induced chaos self-destructs. The LLE values estimated by using the Rosenstein algorithm[33] as shown in a1-f1 are different from those (a1-f1: -0.71, -0.67, -0.65, -0.65, 0.23, 0.18) calculated directly from Eq. 2 in theory. After the series are subsampled to eliminate the periodic skeleton, NL becomes positive in a1 (NL ~ 10%) indicating subtle noise-induced intermittent chaos but LLE remains negative in both a1 and b1.

coin tossing experiment, where the outcome (heads or tails) is *discrete* and *symbolic* rather than dynamic, and remains unchanged (*static*) once the experiment is over. We call this '*discrete-event chaos*' (Table 1). A canonical model of discrete-event chaos is the Smale horseshoe map, a topological construct that emulates the coin tossing experiment. Widely hailed as one of the earliest examples of autonomous deterministic chaos, the Smale horseshoe was actually inspired by the theory of nonautonomous chaotic fluctuations in the periodically-driven van der Pol oscillator [see ref.[41]].

An interesting example demonstrating both nonautonomous deterministic chaos and discrete-event chaos is the following periodically-driven system:

$$x_{n+1} = 2.4\sin(x_n) + T(t) \qquad [1]$$

where $T(t)$ is a square wave switching between 0 and 1.2. The underlying autonomous skeleton (with $T(t)$ constant at 0 or 1.2) exhibits period-2 or period-3 oscillations (Supplementary Fig. S1). However, with judicious choices of the period and duty cycle of $T(t)$, Eq. 1 exhibits vastly different dynamics. In Fig. 1a, the solution simulates the coin tossing experiment with fractal basins of attraction for two antiphase limit cycles representing symbolically the "heads" or "tails" of the discrete-event outcome. In this case, sensitive dependence on initial conditions occurs in the fractal basins of attraction instead of the resultant limit cycles. In contrast, in Fig. 1b the solution converges to a single limit cycle with uniform basin of attraction. As expected, NL = 0 and LLE < 0 in the steady state for both discrete-event chaos (Fig. 1a) and single periodic orbit (Fig. 1b).

However, close examination of Fig. 1a shows that the basins of attraction for discrete-event chaos are not truly "fractal" as many adjacent boundaries have finite widths. Such coarse-grained fractal structure is also characteristic of the coin tossing experiment in real life. Nevertheless, since human hand movements have finite precision according to Fitts' law[42], the coin tossing outcome is always sensitive to the intended initial conditions even though the actual basin boundaries are not truly fractal in a strict mathematical sense. Thus, although chaos theory is by definition infinitesimally precise, its application to real-life problems must take into account the finite precision and randomness of physical, biological and socioeconomic experiments.

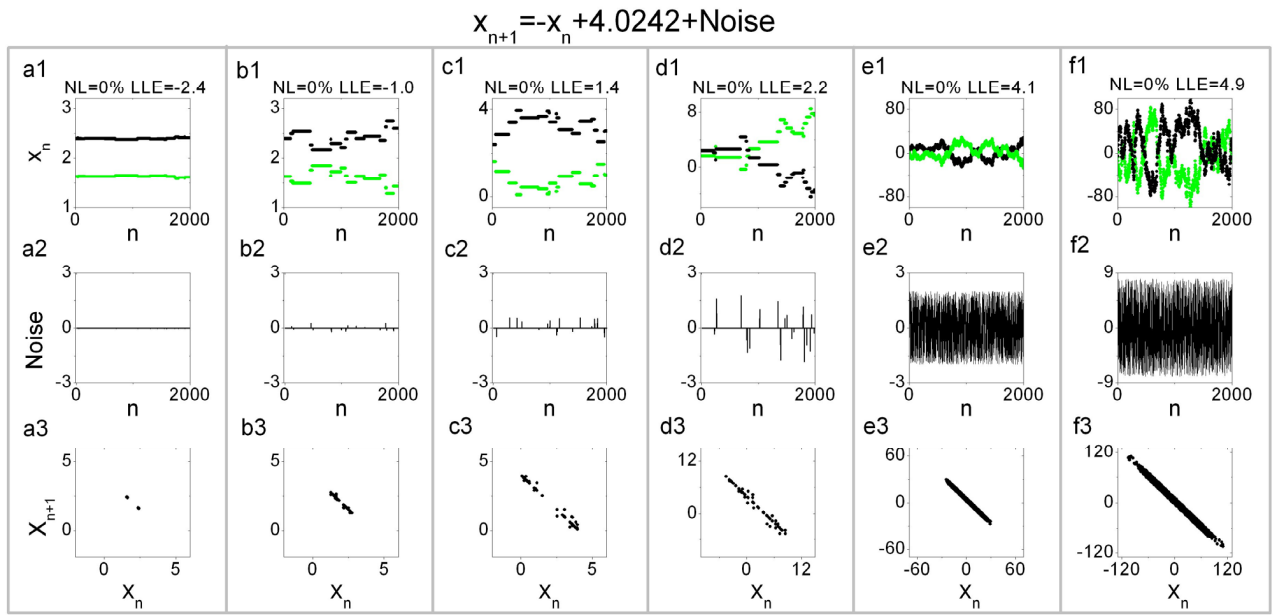

Fig. 3 Examples of linearly correlated colored noise with NL = 0. The linear system (Eq. 3) is driven by the same increasing levels of noise (a2-f2) as in Fig. 2 with initial condition x0=1.628. a1-f1 are the resulting time series and a3-f3 are corresponding first return maps. The LLE's are estimated by using the Rosenstein algorithm33. Results are similar regardless of whether the series are subsampled to eliminate the periodic skeleton.

In Fig. 1c, the square wave $T(t)$ is constant at 1.2 most of the time except resetting to 0 once every 20 steps. Rather than settling in a period-3 orbit as predicted by the autonomous skeleton (Supplementary Fig. S1b), the resulting orbit becomes chaotic (LLE > 0) and much more complex. The nonautonomous deterministic chaos is correctly detected by noise titration (NL > 0).

**Titration of noise-induced nonautonomous chaos**

Equation 1 highlights the danger of attempting to predict nonautonomous dynamics from the autonomous skeleton. If deterministic inputs can induce discrete-event chaos or even aperiodic nonautonomous deterministic chaos in otherwise nonchaotic systems, so too can dynamic noise inputs[15,16]. Consider Eq. 1 with $T(t)$ replaced by a noise input as follows:

$$x_{n+1} = 2.4\sin(x_n) + noise \qquad [2]$$

Equation 2 is a nonlinear stochastic difference equation in the Langevin form. For Gaussian white noise input the time evolution of the probability distribution function of $x_n$ is described by the Fokker-Planck equation[43]. For arbitrary noise inputs a formal general solution cannot be given. Freitas et al.[44] recently referred to Eq. 2 as "deterministic nonchaotic randomly driven dynamics" by ignoring the inherent nonlinear dynamics, but their prediction of NL ≈ 0 based on this assumption was contradicted by their data.

To demonstrate how dynamic noise may interact with nonlinear dynamics to induce nonautonomous chaos in a chimeric deterministic-stochastic state, we subject Eq. 2 to increasing levels of dynamic noise. At low noise levels the steady-state solution displays predominantly period-2 oscillations with NL = 0 and LLE < 0 (Fig. 2a). At an intermediate noise level the period-2 oscillation becomes intermittently chaotic with NL > 0 and a more fractal-like return map (Fig. 2b), while the LLE remains < 0. This is analogous to a class of chaotic fluctuations that are difficult to discern by nonlinear forecasting because of an attendant periodic skeleton[31] [see also Fig. 1 in ref.[32]]. Subsampling the time series in Figs. 2a, 2b to eliminate the periodic skeleton[31] results in NL > 0 while LLE remains < 0 in both cases. Thus, noise titration is highly sensitive in detecting noise-induced intermittent chaos (Table 1).

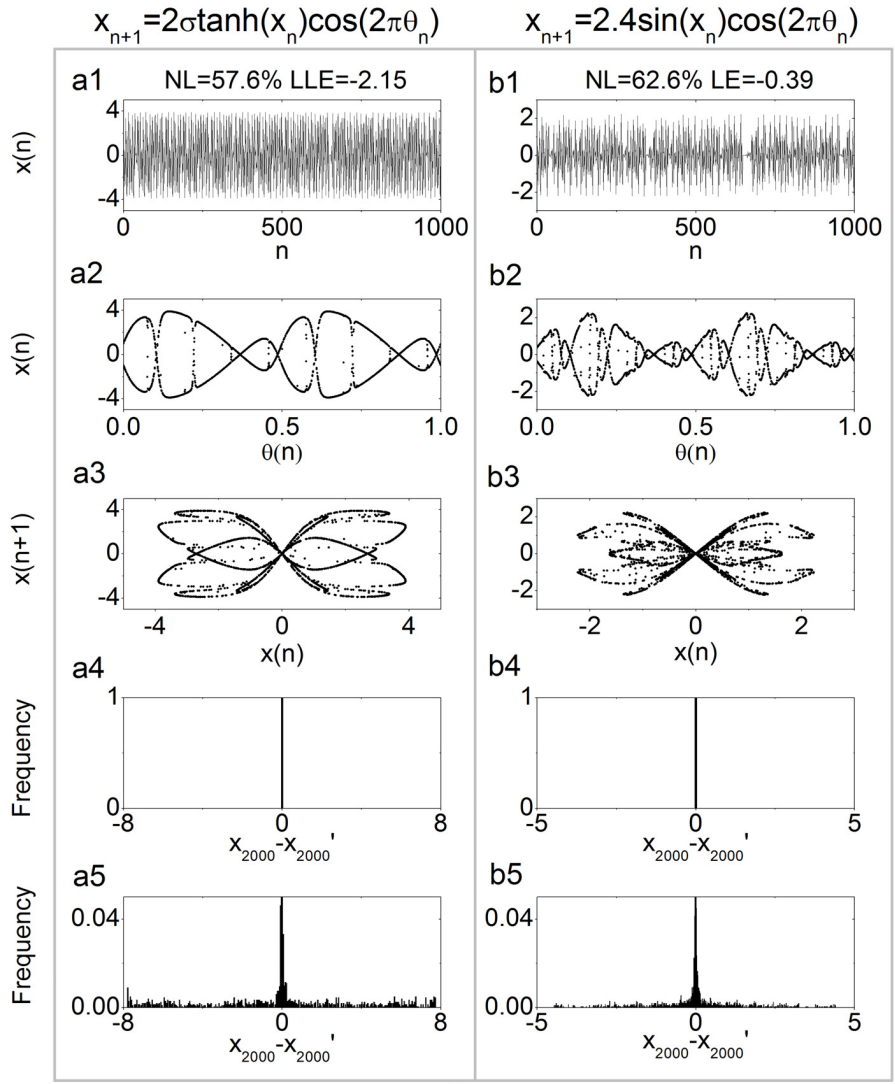

Fig. 4 Li-Yorke chaos in strange nonchaotic attractors. Model a is a classic strange nonchaotic attractor46 (with σ = 2) and Model b is as shown in Eq. 5. Both models are driven by a quasiperiodic input where ω is the golden mean (1.61803399). a1-a3, b1-b3 are corresponding time series xn, fractal attractors and first return maps. a4 and b4 are the histograms of the difference (x2000 – x2000') resulting from two initial conditions (x0, θ0) and (x0', θ0), where θ0 = 0.5, x0' = x0+0.001, and x0 is uniformly random on (-4, 4) for Model a and (-3 3) for Model b. a5 and b5 are the histograms of the difference (x2000 – x2000') resulting from two initial conditions (x0, θ0) and (x0', θ0'), where x0 = x0' = 0.5, θ0' = θ0+0.001, and θ0 is uniformly random on (0, 1) for both models. Both models are Li-Yorke chaotic with NL > 0, even though in both cases the LLE for xn (calculated directly from the equations) is negative and the LLE for θn is zero.

Remarkably, with even higher dynamic noise levels the resultant orbit exhibits increasingly drastic variability ranging from intermittent phase resetting (abrupt phase reversal of the period-2 oscillation, Fig. 2c) and basin hopping (quantum jumps to another periodic orbit and back, Fig. 2d), to becoming full-blown aperiodic altogether (Fig. 2e; Table 1). The increasing unpredictability of the long-term trajectory is also indicated by an increasingly fractal return map with NL becoming increasingly more positive. Such long-term unpredictability can be seen more clearly on longer timescales (Supplementary Fig. S2). In all cases, the resultant noise-induced chaos displays much more complex fluctuations than the periodic skeleton or the noise itself as the orbit is sensitively dependent on the noise input, in that distinct initial values of noise input (not initial values of the periodic skeleton) result in distinct orbits.

With excessive dynamic noise (Fig. 2f) the resultant orbit becomes random with NL = 0 as the orbit cannot be fit by a low-dimensional linear or nonlinear VAR model, yet LLE remains positive in this case. This is consistent with previous mathematical analysis which shows that any nonlinear process (chaotic or otherwise) that is perturbed by dynamic noise may be approximated arbitrarily closely by a linear model of sufficiently high order[45]. Thus, the presence of large dynamic noise may 'annihilate' autonomous deterministic chaos[18,32,34]. Likewise, Fig. 2f shows that noise-induced chaos may 'self-destruct' and annihilate itself when the noise is excessive (Table 1).

In contrast to the noise-induced chaos in Figs. 2b-2e, consider the following surrogate linear system with identical autonomous period-2 oscillations as in Eq. 3:

$$x_{n+1} = -x_n + 4.0242 + noise \qquad [3]$$

Figure 3 shows that Eq. 3 always results in NL = 0 when subjected to the same incremental levels of noise perturbations as with Eq. 2, whereas LLE shows false positives at higher noise levels. The results are similar regardless of whether the original series are subsampled to eliminate the periodic skeleton. Thus LLE is not a reliable index of noise-induced chaos[16-18].

## A generalized theory of chaos

Li and Yorke[25] first defined deterministic chaos mathematically for first-order autonomous systems as follows. Let $f(\cdot)$ be a continuous map with a bounded nonperiodic attractor in the form of an uncountable set $S$ of real numbers. Then $f$ is said to be chaotic if and only if for every $x, y \in S$ with $x \neq y$,

$$\liminf_{n \to \infty} |f^n(x) - f^n(y)| = 0$$
$$\limsup_{n \to \infty} |f^n(x) - f^n(y)| > 0 \qquad [4]$$

Equation 4 stipulates that $f$ is unpredictable as $n \to \infty$; however, an exponential rate of divergence of $f$ at $n = 0$ (as in nonlinear forecasting[9]) is not required. Thus, a system may be chaotic in the Li-Yorke sense even when LLE < 0. This is illustrated by the discrete-event chaos example shown in Fig. 1a but may be extended to aperiodic forms of nonautonomous deterministic chaos. To show this, consider the following nonautonomous system:

$$\theta_{n+1} = \theta_n + \omega \quad \mod(1)$$
$$x_{n+1} = 2.4 \sin(x_n) \cos(2\pi \theta_n) \qquad [5]$$

where $\theta_n$ is a quasiperiodic input. This system is analogous to the classical strange nonchaotic attractor[46,47]. Figure 4 shows that in both cases, the series $x_n$ has a negative LLE yet its long-term value (as $n \to \infty$) is sensitively dependent on the initial value $\theta_0$ of the input instead of $x_o$. The resultant chaotic trajectories (with NL > 0) are much more complex than the limit cycle skeleton or the quasiperiodic input itself. The occult deterministic chaos is also detected by nonlinear forecasting based on the attractor independent of the LLE (Table 1). Other forms of sensitive dependence on initial conditions for strange nonchaotic attractors without imposed conditions on initial rates of divergence have been described in ref.[48].

More generally, for any low-dimensional nonlinear dynamical system $f(\cdot)$ with bounded deterministic input $\xi$, we extend the Li-Yorke definition of chaos by substituting the nonautonomous functions $f(\mathbf{x}, \xi_1)$ and $f(\mathbf{y}, \xi_2)$ into

| | Nonlinear Dynamics | | | | | | Statistical Physics | | | | | |
|---|---|---|---|---|---|---|---|---|---|---|---|---|
| | Autonomous deterministic | | | Nonautonomous deterministic | | | Nonautonomous stochastic | | | | Autonomous stochastic | |
| | Equili-brium | Limit cycles | Intermittent chaos | Aperiodic chaos | Aperiodic chaos | Discrete-event chaos | Strange nonchaotic attractor | Noise-induced chaos | | Noise Annihilation of chaos | Linearly correlated (colored) noise | High-dimensional chaos | White noise |
| | | | | | | | | Intermittent | Aperiodic | | | | |
| LLE | – | – | x | + | + | x | x | x | + | x | x | x | x |
| NF  | – | – | x | + | + | x | + | x | + | – | – | – | – |
| NL  | o | o | + | + | + | x | + | + | + | o | o | o | o |

Table 1. Autonomous and nonautonomous problem domains in nonlinear dynamics and statistical physics
Shadings (dark or light gray) indicate demarcations between: (i) deterministic-stochastic problem domains in current nonlinear dynamics and statistical physics fields; and (ii) autonomous-nonautonomous deterministic problem domains as defined herein. The nonautonomous deterministic chaos and noise-induced (chimeric deterministic-stochastic) chaos domains provide a critical missing link between traditional autonomous deterministic and autonomous stochastic domains. Classes of chaos in shaded areas are as defined in Figs. 1-4. The symbols + (positive), – (negative) and o (zero) denote correct signs of the largest Lyapunov exponent (LLE), nonlinear forecasting (NF) and noise limit (NL) indicating low-dimensional chaos (or the lack thereof) when measurement noise is nil; the symbol x denotes incorrect sign (false +, – or o for low-dimensional chaos). LLE is prone to false + under significant measurement noise; NF is robust to white but not linearly correlated measurement noise, whereas NL is robust to both (see text). Excessive measurement and/or dynamic noise may annihilate the underlying low-dimensional chaos. Note that NL correctly discriminates low-dimensional chaos from noise in both deterministic and stochastic domains, failing (with false o) only in the case of discrete-event chaos with fractal basins of attraction (Fig. 1a).

Eq. 5, where $\xi_1$, $\xi_2$ are realizations of $\xi$ and $\mathbf{x}, \mathbf{y} \in S^k$ are delay embedding vectors of low dimension $k$ with $(\mathbf{x}, \xi_1) \neq (\mathbf{y}, \xi_2)$:

$$\liminf_{n \to \infty} \left| f^n(\mathbf{x}, \xi_1) - f^n(\mathbf{y}, \xi_2) \right| = 0$$
$$\limsup_{n \to \infty} \left| f^n(\mathbf{x}, \xi_1) - f^n(\mathbf{y}, \xi_2) \right| > 0 \qquad [6]$$

Equation 6 gives a necessary and sufficient condition for all classes of deterministic chaos including autonomous and nonautonomous deterministic chaos, discrete-event chaos and strange nonchaotic attractor (Table 1). Again, the condition LLE > 0 is not necessary.

Although Eq. 6 is mathematically precise, its application to real-world problems is confounded by the ubiquitous presence of measurement and/or dynamic noise. The effect of white or linearly correlated measurement noise is readily discriminated by noise titration, as demonstrated previously[34]. In principle, the presence of even infinitesimal dynamic noise (when $\xi$ in Eq. 6 has a stochastic component) may produce a chimeric deterministic-stochastic state that is unpredictable deterministically. However, when dynamic noise effects are negligible the time series could be equally fit by a linear model (Figs. 2a and 3a); hence chaos is not indicated by noise titration until dynamic noise is strong enough to exacerbate the underlying low-dimensional nonlinear dynamics (Figs. 2b-2e and 3b-3e). On the other hand, excessive dynamic noise or measurement noise may annihilate low-dimensional chaos rendering it poorly fit by nonlinear or linear models (Figs. 2f and 3f).

These observations lead to a proposed generalized definition of chaos as *low-dimensional nonlinear dynamics with long-term unpredictability (Eq. 6)* regardless of whether the chaos is autonomous or nonautonomous, deterministic or noise-induced (Table 1). Under this generalized framework the condition NL > 0 gives a sufficient proof of low-dimensional chaos irrespective of the value of LLE, and the value of NL provides a relative measure of the chaotic intensity. Conversely, the condition NL = 0 does not necessarily rule out low-dimensional chaos as noise-titration is not applicable to discrete-event chaos or when chaos is too weak or obfuscated by excessive measurement and/or dynamic noise.

**Discussion**

Classical statistical physics theories treat all complex signals as stochastic irrespective of possible influence of chaotic dynamics. In contrast, nonlinear dynamics methods for detecting deterministic chaos are often confounded by the ubiquitous presence of noise and uncertainty in real life. Transcending this classic deterministic-stochastic chasm, the foregoing theory forges a unification of autonomous and nonautonomous classes of chaos (Table 1) that demonstrate low-dimensional nonlinear dynamics with long-term unpredictability. The most striking implication of this generalized theory is that deterministic or noise-induced chaos does not necessarily imply LLE > 0 (Eqs. 4, 6) with exponential decay of short-term predictability[9]. This is demonstrated by nonautonomous chaos examples such as discrete-event chaos (Fig. 1a), strange nonchaotic attractor (Fig. 4) and noise-induced chaos in a chimeric deterministic-stochastic state (Figs. 2b,c), where the resultant orbits show sensitive dependence on initial values in the fractal basins of attraction or in the inputs instead of initial values within the attractors. In such cases, interaction of nonlinear dynamics with extrinsic deterministic or stochastic forcing results in chaotic fluctuations that are much more complex than the autonomous skeleton or the inputs themselves.

Under this overarching framework, our results show that noise titration provides a specific, sensitive, and robust (criteria *i-iv*) assay for nearly all classes of low-dimensional chaos (except discrete-event chaos; Table 1) many of which have escaped detection by classical nonlinear dynamics techniques. These findings illuminate previous confusions about the limitations of LLE as a signature of deterministic chaos[16-18,46,47] vis-à-vis the power of the noise titration technique as a sufficient test of both autonomous and nonautonomous chaos[44,49]. The robust discrimination power of noise titration in assaying low-dimensional chaos in the presence of measurement noise allows real-world data to be reliably screened for amenability to finite-state systems modeling, leaving only intractable chaos (high-dimensional chaos and noise-annihilation of low-dimensional chaos; Table 1) as ultimate stochastic events that warrant statistical analyses.

Since noise titration is a statistical hypothesis testing procedure, its accuracy is contingent on the power of the chosen nonlinear discrimination model and test statistic, as well as the quality and quantity of the test data. Noise titration with the VAR model has been shown[34] to accurately discern chaotic dynamics in short, noisy data with dimensions of up to 20. With the evolution of increasing computing power and more sensitive nonlinear discrimination models and more robust test statistics (especially for nonstationary or non-Gaussian error distributions), the frontier of nonlinear dynamics problem domain will likely continue to advance toward even higher-dimensional (more complex) systems in future.

Presently, no nonlinear dynamics or statistical physics techniques are available to distinguish deterministic and noise-induced chaos particularly when the data is marred by significant measurement noise. Distinguishing measurement noise from dynamic noise effects in nonlinear time series analysis is a challenging (perhaps intractable) problem except in special cases[50]. Nevertheless, inasmuch as low-dimensional chaos can be discerned by NL > 0 indicating that the underlying nonlinear mechanisms are amenable to modeling and analysis, the precise classification of the chaos (Table 1) is unimportant. Indeed, low-dimensional nonlinear dynamics is also ascertained once nonlinear limit cycles distinct from linear oscillations (Fig. 3) are discerned, regardless of whether any attendant irregularities to the periodic orbit are classified as deterministic chaos[31,32], noise-induced chaos (Fig. 2b) or sheer 'residual noise' to the dominant periodicity. Such fine details should be best resolved by analyzing the changes in the pseuo-periodic or chaotic components following specific system perturbations, such as between normal and abnormal states or after judicious experimental maneuvers.

## Acknowledgments

CL was supported by the State Scholarship Fund awarded by the China Scholarship Council. This work was supported in part by U.S. National Institutes of Health grants HL079503, HL075014 and HL072849 (CSP), Shanghai Leading Academic Discipline Project, No B112 and National Natural Science Foundation of China grant No. 30370353 (GQW).

# Supplementary Information

## METHODS

***Noise titration method***. In this method[34], nonlinear determinism in a time series is first identified[32,35] by comparing a linear model and a polynomial autoregressive model (Eq. 7) with varying memory order ($\kappa$) and nonlinear degree ($d$) to optimally predict the data based on the Akaike information criterion $C(r)$ (Eq. 8):

$$y_n^{nonlin} = a_0 + a_1 y_{n-1} + a_2 y_{n-2} + \cdots + a_\kappa y_{n-\kappa} + a_{\kappa+1} y_{n-1}^2$$
$$+ a_{\kappa+2} y_{n-1} y_{n-2} + \cdots + a_{M-1} y_{n-\kappa} + \varepsilon(\kappa, d)$$

$$C(r) = \log \varepsilon(r) + r/N$$

In the above, $\varepsilon$ is modeling error; $M$ is total number of polynomial terms in Eq. 7; $r$ is number of leading polynomial terms in Eq. 7 ($1 < r < M$) used in the computation of $C(r)$; $N$ is total number of data points in the series. The parameters $a_m$ in Eq. 7 are recursively estimated. The null hypothesis — a stochastic time series with linear dynamics — is rejected if the best nonlinear model provided a significantly better fit to the data than the best linear model using parametric (F-test) or nonparametric (Mann-Whitney-Wilcoxon) statistics at the 1% significance level.

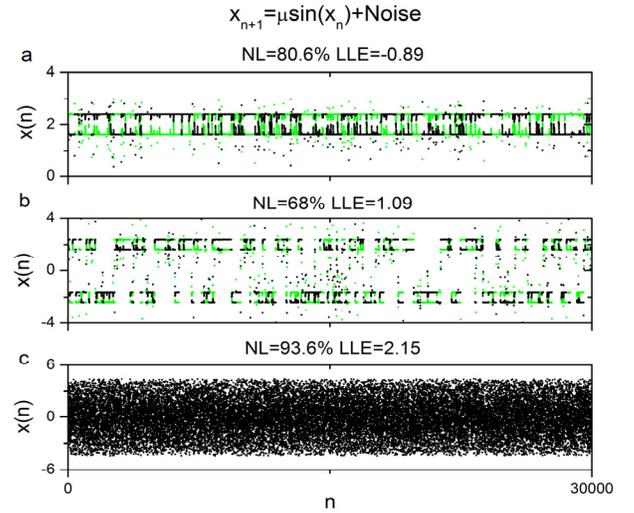

**Fig. S2.** Noise-induced chaos showing long-term unpredictivity as predicted by NL > 0. Results are same as those in Figs. 2c1-2e1 but plotted for 30,000 points. The LLE's were estimated by using the Rosenstein algorithm[33].

Once nonlinear determinism is indicated, white (or linearly correlated) noise of increasing standard deviations is added to the data until nonlinearity could no longer be detected, i.e., the nonlinearity is 'neutralized'. The noise limit (NL) for a specific noise function is calculated as the percent of signal power added as noise to 'titrate' the data to the point of neutrality. Typically, an average NL value is obtained by repeating the titration procedure 5-10 times. Under this scheme, chaos is indicated by NL > 0, and the value of NL provides a relative measure of chaos intensity[34]. Conversely, if NL = 0, then it may be inferred that the series either is not chaotic or the chaotic component is already neutralized by the background noise (noise floor) in the data.

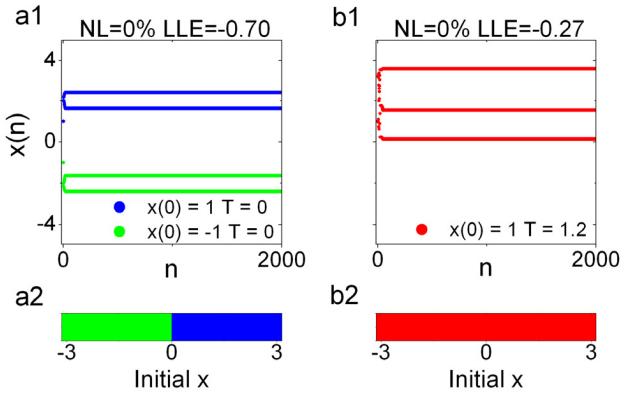

**Fig. S1.** Characteristics of the map $x_{n+1} = 2.4 \sin(x_n) + T$ (autonomous part of Eq. 1) where $T$ is a constant. In *a1*, when $T = 0$, the steady-state solutions comprise two symmetric period-2 antiphase orbits (blue: 2.39606595 \to 1.62806141\...; green: -2.39606595 \to -1.62806141\...) depending on initial conditions. The basins of attraction to these orbits are disjoint with distinct positive or negative basins and a singularity at 0. In *a2*, initial conditions on ($-\pi, \pi$) that eventually result in positive oscillations are marked in blue; those that result in negative oscillations are green. In *b1*, when $T = 1.2$, the system exhibits stable period-3 oscillations (red: 3.5986321 \to 0.140895963 \to 1.5370326...) for all initial values. In *b2*, all initial conditions on ($-\pi, \pi$) that result in this steady-state solution are marked in red. Noise limit (NL) is zero for all periodic solutions. The largest Lyapunov exponent (LLE < 0) was calculated directly from the equation.

13